\documentclass[preprints,article,accept,pdftex,moreauthors]{Definitions/mdpi} 
\firstpage{1} 
\pubvolume{1}
\issuenum{1}
\articlenumber{0}
\pubyear{2022}
\copyrightyear{2022}
\datereceived{} 
\dateaccepted{} 
\datepublished{} 
\hreflink{https://doi.org/} 

\Title{Utilizing geospatial data for assessing energy security: Mapping small solar home systems using unmanned aerial vehicles and deep learning}

\TitleCitation{Title}

\Author{Simiao Ren $^{1}$\orcidR{}, Jordan Malof $^{1}$\orcidM{}, T. Robert Fetter$^{1}$\orcidF{}, Robert Beach$^{2}$\orcidB{}, Jay Rineer$^{2}$\orcidJ{} and Kyle Bradbury $^{1}$*\orcidK{}}

\AuthorNames{Simiao Ren , Jordan Malof, T. Robert Fetter, Robert Beach, Jay Rineer and Kyle Bradbury}

\AuthorCitation{Ren, S.; Malof, J.; Fetter, T. R.; Beach, R.; Rineer, J.; Bradbury, K}

\address{%
$^{1}$ \quad Duke University, Durham, NC, 27705, USA; simiao.ren/jordan.malof/rob.fetter/kyle.bradbury@duke.edu\\
$^{2}$ \quad RTI International, Research Triangle Park, NC 27709, USA; rbeach/jrin@rti.org}

\corres{Correspondence: kyle.bradbury@duke.edu}

\abstract{Solar home systems (SHS), a cost-effective solution for rural communities far from the grid in developing countries, are small solar panels and associated equipment that provides power to a single household. A crucial resource for targeting further investment of public and private resources, as well as tracking the progress of universal electrification goals, is shared access to high-quality data on individual SHS installations including information such as location and power capacity. Though recent studies utilizing satellite imagery and machine learning to detect solar panels have emerged, they struggle to accurately locate many SHS due to limited image resolution (some small solar panels only occupy several pixels in satellite imagery). In this work, we explore the viability and cost-performance tradeoff of using automatic SHS detection on unmanned aerial vehicle (UAV) imagery as an alternative to satellite imagery. More specifically, we explore three questions: (i) what is the detection performance of SHS using drone imagery; (ii) how expensive is the drone data collection, compared to satellite imagery; and (iii) how well does drone-based SHS detection perform in real-world scenarios. To examine these questions, we collect and publicly-release a dataset of high-resolution drone imagery encompassing SHS imaged under a variety of real-world conditions and use this dataset and a dataset of imagery from Rwanda to evaluate the capabilities of deep learning models to recognize SHS, including those that are too small to be reliably recognized in satellite imagery. The results suggest that UAV imagery may be a viable alternative to identify very small SHS from perspectives of both detection accuracy and financial costs of data collection. UAV-based data collection may be a practical option for supporting electricity access planning strategies for achieving sustainable development goals and for monitoring the progress towards those goals.   }


\dataset{Main drone dataset: \url{10.6084/m9.figshare.18093890}; Rwanda PV annotations: \url{10.6084/m9.figshare.18094043};}

\begin{document}

Energy access is a global issue. Sustainable development goal (SDG) 7.1, outlined by United Nations \cite{UNSDG}, proposed to achieve "access to affordable, reliable, sustainable, and modern energy for all" by 2030. However, as the latest progress report in 2020 \cite{martin_sustainable_nodate} pointed out, extrapolating from the current progress towards electrification, there would still be 620 million people without basic access to electricity by 2030. Although the great majority of the global population has grid-connected electricity access (only 35 million people worldwide relied solely on off-grid solutions in 2018 \cite{off-grid-stats}) the high costs of traversing challenging topographies make grid extension to remote, rural communities less affordable or efficient  \cite{bisaga2021mapping}.

In attempting to reach SDG 7.1, solar home systems (SHS) have become a promising solution as an alternative to grid extension \cite{bandi2020touching}. SHS are solar-photovoltaic-based systems that provide electricity to individual homes. Compared to solar farms where large solar arrays are installed and generated power is transmitted and regulated by large power grids, a SHS (a solar panel and associated equipment that provides power to a single household) is typically a small solar panel fixed to the top of the household roof and not connected to the electric grid. Off-grid solar systems are often cost-effective solutions for rural communities far from the grid.

A crucial resource for targeting further investment of public and private resources, as well as tracking the progress of universal electrification goals, is shared access to high-quality data on individual SHS installations including information such as location and capacity. With this information, decision-makers can make more informed decisions about electrification options, such as grid extensions, mini/micro grids, and stand-alone systems \cite{watson2019renewable}. Despite the importance of such data, unfortunately this information is often of limited availability or only accessible by expensive and time-consuming surveys \cite{malof2015automatic} or incomplete self reports \cite{Castello2019}. 

To address this widespread obstacle to energy access tracking and planning \cite{bhatia2015beyond}, previous studies have recognized that SHS are visible in overhead imagery and when remote sensing data sources are combined with with machine learning tools, it may create a suitable approach to data collection for some SHS. Utilizing high-resolution satellite images, it was shown that medium to large installations of solar panels can be identified with high accuracy \cite{malof2016automatic, yu2018deepsolar, Malof2016,kruitwagen2021global}. However, significantly smaller sizes of SHS are used in rural areas of low- and middle-income countries. The majority of the global population using off-grid solutions use SHS under 50W  \cite{off-grid-stats}, which typically occupies an area of $0.3m^2$\cite{solar_panel_size}. This is a challenge even at the highest resolution of commercially available satellite imagery (typically around 0.3m/pixel \cite{liang_chapter_2020}). This is a major limitation of the collection of location and capacity data for very small SHS. Therefore, unmanned aerial vehicles (UAV), offer a higher-resolution alternative to satellite platforms and are a potential alternative solution to filling this information gap. A more detailed discussion of the existing literature on solar panel detection using remote sensing data is presented in Section \ref{related_work}.

\subsection{Contributions of this work} 
\label{sec:research_questions_contribution}
In this paper, we critically analyze the viability of using drone-based aerial imagery to detect common, but physically small, energy infrastructure as an alternative to satellite imagery or manual surveys from both an assessment accuracy and cost effectiveness perspective. Our contributions closely follow the three research questions that we propose to answer: (i) what is the detection performance of SHS using UAV imagery; (ii) how expensive is the drone data collection, compared to satellite imagery; and (iii) how well does drone-based SHS detection perform in real-world scenarios. 

For investigating (i) the detection performance of SHS using UAV imagery, as there exists no public dataset to directly answer this question, we develop a drone-based solar panel detection dataset covering various flying altitude and flying speed and train a deep learning detection algorithm to detect the SHS. We then evaluate the performance and robustness of this algorithm. For research question (ii) regarding the comparative cost of UAV based data collection, we conduct a cost/benefit analysis of UAV-based SHS data collection and compare it to common alternatives such as satellite imagery and aerial photography. For research question (iii) regarding the performance of UAV-based data analysis for SHS assessment under real-world conditions, we annotated and evaluated SHS detection performance on UAV imagery from Rwanda to quantify the performance of UAV imagery in a real-world setting.

The main contributions of this work are summarized as follows:
\begin{itemize}
    \item \textit{The first publicly available dataset of UAV imagery of very small (less than 100 Watts) SHS (Section \ref{sec:data})}.  We collected, annotated, and openly shared the first UAV-based very small solar panel dataset with precise ground sampling distance and flight altitude. The dataset contains 423 images, 60 videos and 2019 annotated solar panel instances. The dataset contains annotations for training object detection or segmentation models. 

    \item \textit{Evaluating the robustness and detection performance of deep learning object detection for solar PV UAV data (Section \ref{sec:exp1})}.  We evaluate the performance of SHS detection performance with a U-Net architecture with a pre-trained ResNet50 backbone. We controlled for the data collection resolution (or 1/altitude): sampling every 10m of altitude across an interval from 50m-120m.  We controlled for the dimension of panel size by using 5 diverse solar PV panel sizes

    \item \textit{Cost/benefit analysis of UAV- and satellite-based solar PV mapping(Section \ref{sec:exp2})}. We estimate a cost-performance curve for comparing remote sensing based data collection for both UAV and satellite systems for direct comparison. We demonstrate that using the highest resolution satellite imagery currently available, very small SHS are hardly detectable; thus even the highest-resolution commercially available satellite imagery does not present a viable solution for assessment of very small (less than 100 Watt) solar panel deployments.
    
    \item \textit{Case study in Rwanda illustrating the potential of drone-based solar panel detection for very small SHS installations(Section \ref{sec:exp3})} By applying our models to drone data collected in the field in Rwanda, we demonstrate an example of the practical performance of using UAV imagery for solar panel detection. Comparing the results to our experiments with data collected under controlled conditions, we identified the two largest obstacles to achieving improved performance are the resolution of the imagery and the diversity of the training data.
\end{itemize}



\section{Related work} \label{related_work}
Larger solar PV arrays, especially as compared to very small SHS, have been shown to be detectable in satellite imagery. Recent work demonstrates the potential of automatically mapping Solar PV arrays using remote sensing and machine learning from individual household SHS (5-10kW \cite{malof2015automatic, Yuan2016Large,Bradbury2016, malof2016automatic, yu2018deepsolar}) to utility-scale ($>$10MW \cite{kruitwagen2021global, ishii2016detection}) installations, while no past studies have focused on smaller solar arrays that are being deployed for those transitioning to electricity access for the first time, which could be 100W or less - an order of magnitude smaller than all past studies. 

The earliest work on solar panel segmentation used traditional machine learning methods like support vector machines (SVM), decision trees and random forests, with manually-engineered features like mean, variance, textons, and local color statistics of pixels \cite{malof2015automatic,malof2016automatic,Malof2016b}. Yuan et al. \cite{Yuan2016Large} applied deep neural networks (DNN) for solar panel detection from aerial imagery. Advances in convolutional neural networks (CNNs) \cite{krizhevsky2012imagenet,long2015fully} in large scale image datasets like ImageNet \cite{deng2009imagenet} have also propelled the field of solar panel segmentation (i.e. pixel-wise classification) forward. Convolutional networks for image classification were the first type of CNNs to be used for a coarse form of solar panel segmentation by \cite{Malof2016,Yuan2016,Malof2019}. True segmentation CNNs for solar PV identification soon followed including SegNet \cite{Camilo2018} and activation map based methods \cite{yu2018deepsolar}. U-net structures \cite{ronneberger2015u} were also quickly adopted in this field further increasing model detection performance \cite{Castello2019, Hou2019}.


UAVs have been used for solar PV monitoring and management on solar farms. However, this has typically been limited to situations where the locations of the solar PV is already known such as solar panel monitoring in large solar farms. UAV-based solar array segmentation has been used for inspecting the arrays of solar farms, using texture features and a clustering algorithm for the analysis \cite{Zhang2017}.  Other research on drone-based solar panel detection used a thermal camera to identify potential defects in solar arrays at the site \cite{ismail2019autonomous, vega2020solar}.  There have also been techniques for monitoring solar PV farms using satellite imagery, such as tracking particulate matter deposition and effects on generation efficiency \cite{zheng2020estimating}; however, this is the exception since solar PV condition monitoring typically requires UAV data due to high image resolution requirements. In these settings both thermal and optical cameras have been used. Thermal cameras are used in this setting to identify  abnormal temperature profiles that may indicate malfunctioning solar arrays \cite{Pierdicca2018, Xie2020, Herraiz2019}. Optical UAV imagery has also been used to identify damaged or dust covered solar cells at solar farms \cite{Li2019a,Ding2019,Li2018, Li2020,Hanafy2019}. None of the these UAV-based studies released their imagery datasets.

\section{The SHS Drone Imagery Dataset} \label{sec:data}

Drone datasets containing annotated images of solar panels, and especially very small ( $\leq$ 100W) solar panels do not exist to our knowledge. We summarize the most relevant publicly-available UAV-based datasets in the appendix; however, none of these datasets contain solar panel annotations, making them unsuitable for training automated techniques to identify solar PV cells in UAV imagery.

Since no existing public data were available for our study, we collected a new dataset\footnote{Figshare DOI: \url{10.6084/m9.figshare.18093890}} specifically for exploring the performance of UAVs (a.k.a. drones) under controlled conditions for identifying very small solar PV. We describe our dataset, test sites, and equipment below. Below are our data design considerations:
\begin{enumerate}
    \item Adequate ground sampling distance (GSD) range and granularity. Due to variations in factors such as hardware and elevation change, the GSD of drone imagery can vary significantly in practice. Therefore we want our dataset to contain imagery with a range of image GSDs that are sufficient to represent a variety of real-world conditions, as well as detect SHS. 
    \item Diverse and representative solar panels. As solar panels can have different configurations affecting the visual appearance (polycrystal or monocrystal, size, aspect ratio), we chose our solar panels carefully so that they form a diverse and representative (in terms of power capacity) set of actual solar panels that would be deployed in developing countries.
    \item Fixed camera angle of 90 degrees and different flying speeds: To investigate the robustness of solar panel detection as well as data collection cost (that is correlated with flying speed), we want our dataset to have more than one flight speed.
\end{enumerate}


\begin{table}[h!]
    \centering
    \label{Tbl:dataset}
    \begin{tabular}{ccccccccc}
    \toprule
    Altitude & GSD & \# Img & \# Vid & \# Annotated PV \\
    \midrule
    50m & 1.7cm & 58 & 6 & 248\\
    60m & 2.1cm & 63 & 7 & 289\\
    70m & 2.5cm & 47 & 8 & 227\\
    80m & 2.8cm & 60 & 8 & 295 \\
    90m & 3.2cm & 44& 8 & 214\\
    100m &3.5cm & 47 & 9 & 230\\
    110m &3.9cm & 56 & 4 & 278\\
    120m &4.3cm & 48 & 10& 238\\
    \bottomrule
    \end{tabular}
    \caption{Details of the dataset collected. Altitude and corresponding GSD are listed. Img: Images. Vid: Videos. In total there are 423 images, 60 videos and 2019 annotated solar panels.}
\end{table}
\subsection{Data collection process}
\textbf{UAV}. A DJI mini 2 was used to collect all our data except for the Rwanda case study (in which we used previously published data\cite{chew2020deep, RTI_dataset}). The DJI mini 2 UAV was selected due to its flexibility in operational mode, high camera quality, and low cost. It has 3 modes of flight, with a maximum speed of 16m/s. It has the capability to hover nearly stationarily, resulting in high image quality in low wind conditions. It uses 1/2.3'' CMOS sensor (4k x 3k pixels) with a wide range of possible shutter speeds. It shoots 4k video (3840x2160) up to 30 fps. We fixed the gimbal (angle of camera) to be vertically downward (90 degrees).

\textbf{Solar panels}. We purchased 5 solar panels in total with various physical sizes and PV materials to be representative of the types of SHS that are typically installed for offgrid solar projects in communities transitioning to electricity and include the low end of feasible use (around 10W). More detailed specifications of our solar panels are presented in the appendix.

\textbf{Flight plan}. Imagery data were collected in the flight zone within Duke Forest in Durham, NC. Due to regulations of the federal aviation agency (FAA) under part 107, we were allowed to occupy class G airspace that includes altitudes lower than 120 meters. Therefore, the height of all the data collected was capped at 120 meters. Also, to prevent collisions with trees, we set the lower limit of our flight height to be 50m.

\textbf{Annotation}. The data were manually annotated with polygons drawn around the solar panel by a human annotator. These polygons were converted to pixel maps for network training. All stationary images were labelled and a small subset of the videos were annotated.

\begin{figure}[h!]
    \centering
    \includegraphics[width=\textwidth]{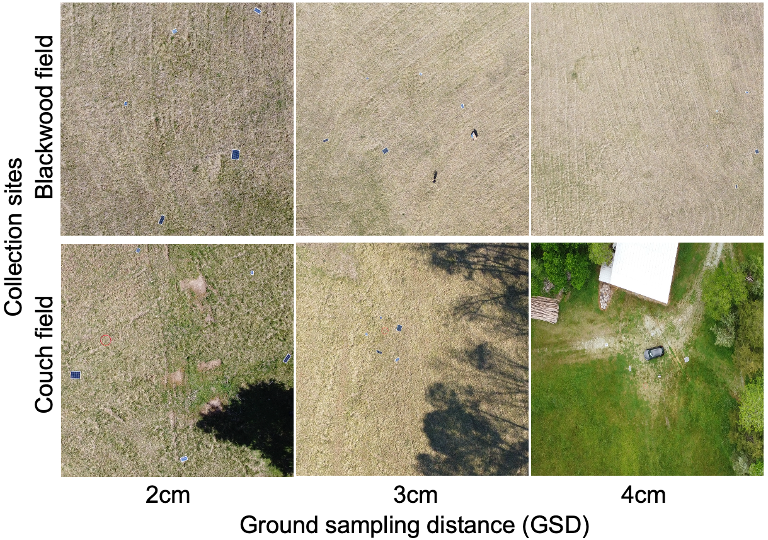}
    \caption{Example imagery in our dataset collected. (a)-(c) are imagery taken in Blackwood Field. (d)-(f) are imagery taken in Couch Field. (a) and (d) have GSD of ~2cm, (b) and (e) has GSD of ~3cm, (c) and (f) has GSD of ~4 cm}
\label{Fig:Sample_imagery_collected}
\end{figure}

We present examples of our collected dataset here for reference in Fig \ref{Fig:Sample_imagery_collected}.

\section{Post-processing and metrics}

\begin{figure*}[h!]
	\begin{adjustwidth}{-\extralength}{0cm}
    \centering
    \includegraphics[width=1.4\textwidth]{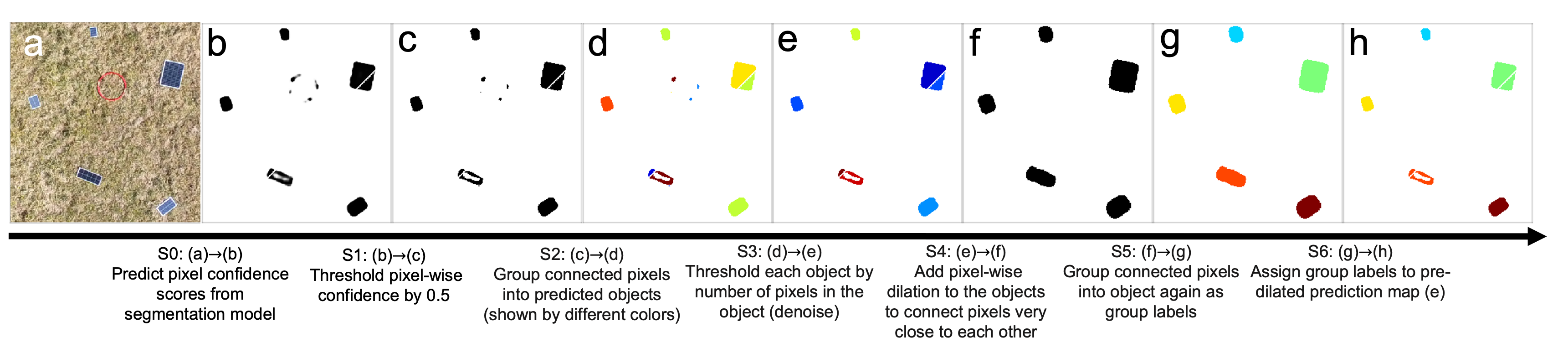}
    \end{adjustwidth}
    \caption{Post-processing flow schematic chart. (a) Original RGB drone imagery.  The post processing step takes in the prediction confidence map (b) from the model output and generates candidate objects through thresholding, grouping, and dilating. Step 1 (S1) thresholds the confidence at 0.5, eliminating the least-confident detections. Step 2 (S2) matches connected pixels into groups of pixels (groups shown in different colors). Step 3 (S3) eliminates the groups of pixels that are too small (likely noise). Until this point, some pixels that corresponds to the same solar panel appear disconnected and therefore belong to different groups. Steps 4 and 5 addresses this issue by dilating the proposal pixels (S4) and grouping them (S5). To ensure the dilation does not change the overall area of prediction, we assign the groups upon the pre-dilation map but with the dilated grouping by label point-wise multiplication. }
    \label{Fig:post_precessing}
\end{figure*}

For the evaluation metrics, we use standard object-wise detection precision and recall, and summary metrics, as described below. As our models are U-Net-based segmentation models \footnote{All our code can be found at \url{https://github.com/BensonRen/Drone_based_solar_PV_detection}} that produce pixel-wise confidence scores, we aggregate the pixel-wise predictions into objects based on thresholding the pixel-wise confidence scores into binary images, grouping those binary images into groups of pixels (objects), and morphologically dilating those objects into groups of neighboring objects, a process illustrated in the Fig \ref{Fig:post_precessing}. The resulting set of objects and corresponding confidence scores then serve as an input to a scoring function that compares each detected object to ground truth to determine whether they are a true positive, false positive, or false negative object, (illustrated in appendix Fig \ref{Fig:Scoring}). We consider a predicted object to be a true positive if its intersection over union (IoU) with a ground truth object is 0.2 or greater. Intersection over union computes ratio of the intersection of the area between two objects with the union of the area included in the two objects.

While all PR-curves can be found in the appendix, we present summary statistics including the maximum $F_1$ score and average precision (AP). $F_1$ is the harmonic mean of precision and recall and is frequently used as a measure of the accuracy of the detection/classification where an $F_1$ score of 1 implies perfect precision and perfect recall. Average precision is a summary statistic representing the area under the PR-curve. An average precision of 1 signifies perfect precision and recall as well. They are defined as follows:

\begin{align} 
    {F_1}_{max} &= \max_{\tau} 2*\frac{P(\tau) * R(\tau)}{P(\tau) + R(\tau)}\\
    AP &= \int_{0}^{R_{max}} P*dr  \label{Eq:F1}
\end{align}

Here, P is Precision, R is Recall, $R_{max}$ is the maximum recall that the model can achieve (usually less than 1 due to object-wise and IoU limitations), $\tau$ is the operation confidence threshold for object-wise scoring that we sweep to produce the PR curve. The $F_1$ score uses the harmonic mean to penalize lower values in either precision or recall. For example, if a detector achieves 90\% precision and 10\% recall, the $F_1$ score is 18\% (a simple mean would yield 50\%).

\section{Experiment \#1: Solar panel detection performance using UAV imagery } \label{sec:exp1}
As we describe the three experiments included in this study, recall that we are investigating three questions: (i) what is the detection performance of SHS using UAV imagery; (ii) how expensive is the drone data collection, compared to satellite imagery; and (iii) how can drone-based SHS detection be applied in real-world scenarios. This section investigates the first (i) of these three questions, and subsequent sections explore the remaining two.

The primary goal of this first set of experiments is to estimate the achievable accuracy of small SHS detection models when they are applied to UAV imagery. We also evaluate how their accuracy varies with respect to two operational settings: the speed of the UAV, and the ground sampling distance (or resolution) of the UAV's imagery. We hypothesize that these two factors will have a significant impact on the achievable accuracy of recognition techniques. Low image resolution can reduce detection accuracy and higher UAV speed can result in motion blur in the imagery, which can also negatively impact detection accuracy. These two factors are likely to vary in practice, making them an important consideration for practitioners and researchers who wish to employ automated recognition models.  

To estimate the accuracy of contemporary PV detection models on UAV imagery, we applied a proven object detection model called U-Net \cite{ronneberger2015u} to the UAV Solar PV dataset we created for this experiment.  The U-Net has been employed in many recent studies involving object recognition in overhead imagery, including solar PV mapping\cite{hou2019solarnet}. Due to the relatively small size of our UAV dataset, we begin by further training our ImageNet-pretrained U-Net model on a \textit{satellite} imagery solar PV panel dataset \cite{Malof2019}. We then \textit{fine-tune} the model on our UAV data (more details about this in the appendix). We evaluated the performance of this fine-tuned detection model as we varied both the resolution of the imagery and the flight speed at which the imagery was collected.  To evaluate the model's performance, we split our data into separate training, validation, and testing sets, which is a widely-used approach to obtain unbiased generalization performance estimates of machine learning models \cite{bishop2006pattern}. Full details of the training and data handling procedure can be found in the Appendix.  




\subsection{Detection performance comparison over imagery resolution} \label{sec:simulated_sate}

To investigate the performance as resolution varies, we train and test image detection models multiple times, varying the resolution of the image for each scenario. Note that we define image \textit{resolution} as the GSD of the UAV imagery (other optics-focused definitions exist \cite{orych2015review}, but are less relevant for this study).  

To evaluate the performance change due to changes in the resolution of the imagery, we fly our drone at various heights from 50m to 120m (the maximum legally allowed height in our jurisdiction), resulting in an image resolution range of 1.6cm to 4.5cm per pixel. We divide our flying height (altitude) into 20m-interval groups for which we evaluate performance. We trained and tested each possible image resolution to evaluate the performance impact of flying at different altitudes (resulting in varying image resolutions).

Additionally, we compare the performance across UAV image resolution to common satellite imagery resolutions, which are an potential alternative approach to collecting data on very small scale solar PV. We simulate the highest commercially available satellite imagery resolution (30cm) as well as high resolution aerial imagery resolution (7.5cm and 15cm) to provide a thorough comparison to UAV performance. To simulate these alternative image resolutions, by resizing the high-resolution UAV imagery to the desired lower resolution \footnote{details in appendix \ref{apx:resizing} }.

\subsubsection{Result} \label{sec:simulated_sate_result}

Even at the highest flying altitude, as shown in Figure \ref{Fig:cost_compare}(b), UAV-based solar PV detection performance had an Average Precision of 0.94 (at 4.5cm ground sampling distance). While this high level of performance is reasonable given that the UAV resolution is high enough high to capture multiple pixels of solar panel content for each SHS, these results massively outperformed satellite imagery resolutions. As the resolution decreases from our UAV imagery to satellite imagery, the detection performance drops monotonically. At 7.5cm (high-resolution aerial imagery), performance remains high. However, at a resolution of 15cm and above, the detection capability of our neural network degrades heavily with only about half of the panels being detected (the maximum recall is 0.6) and at the typical highest satellite resolution (30cm), the performance drops to near zero. Therefore, we conclude that at the best commercially available satellite resolution (30cm), very small SHS PV panels are not feasible to detect. UAVs offer a feasible alternative for small solar PV detection as does high resolution aerial imagery.

\subsection{Detection performance with respect to resolution mismatch}
As demonstrated in the previous section, very small solar PV detection performance depends greatly on resolution, but across UAV resolutions (1.7cm to 4.3cm per pixel in our data collection), performance was less variable. In practice, the resolution may change for multiple reasons. Elevation of the terrain may change, the slope of the landscape and angle of the image may change over the target area and we may fly UAVs at varying heights for data collection (also possible if collected by different operators). Therefore the resolution may not always be identical. In this section, we investigate the impact of mismatch between training image resolution and test image resolution. 

To accomplish this, we divide our flying height (altitude) into 20m-intervals groups for which we evaluate performance. We trained and tested all the possible pairwise combinations of imagery resolution to evaluate the performance impact of a mismatch between training and testing altitude.

\subsubsection{Results} \label{Sec:E1_cross_valid}
\begin{figure}[h!]
    \centering
    \includegraphics[width=0.8\textwidth]{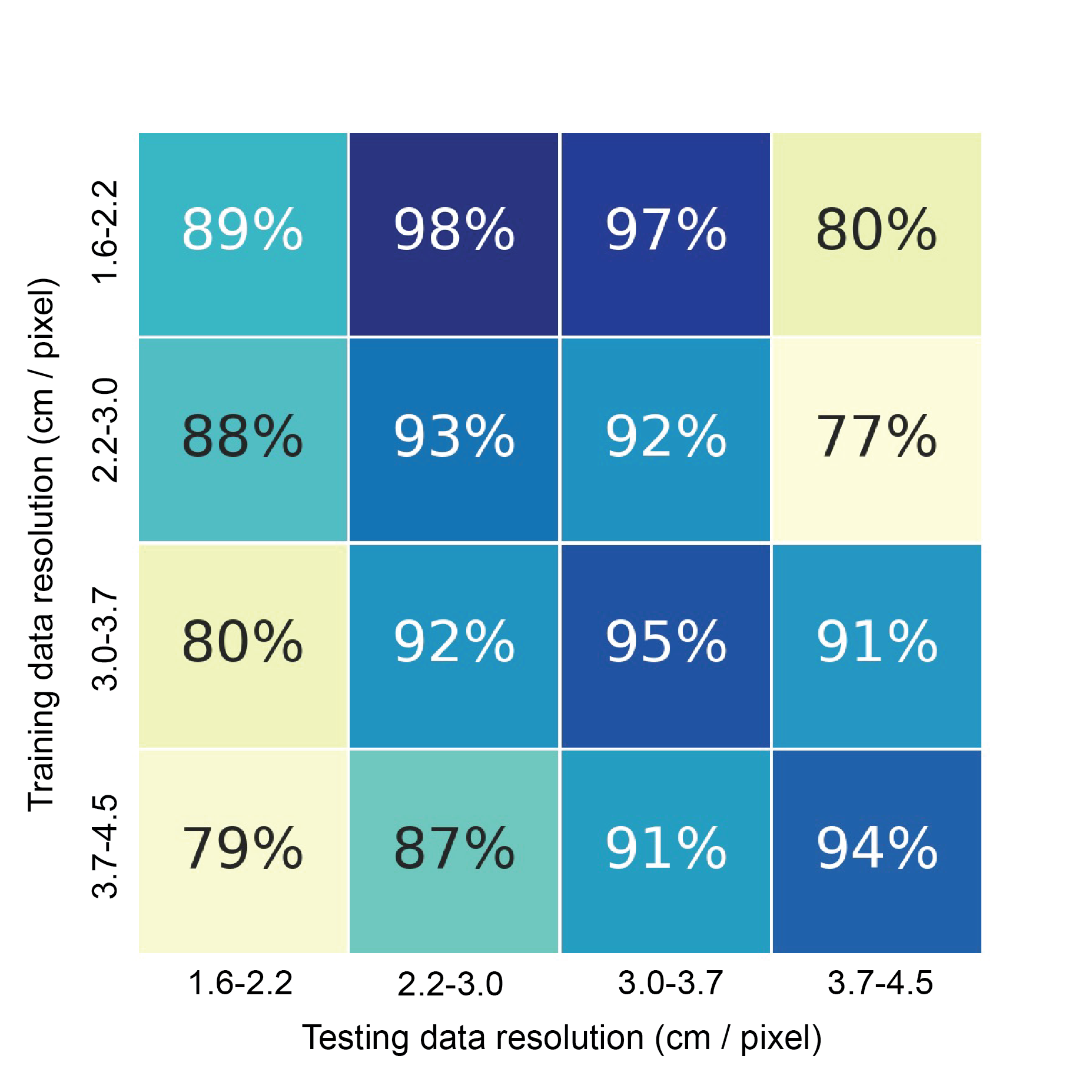}
    \caption{Average Precision score for the training and testing on various resolutions. The rows are the same training resolution and the columns are the testing resolutions. The bins are grouped so that they corresponds to 20m interval in flight height}
\label{Fig:E1_AP}
\end{figure}

The results are shown in the Fig \ref{Fig:E1_AP} where we can see the algorithm's performance with different training and testing image resolution pairings. When the training resolution and test resolution matched, the average precision was ~90\%. This can be seen in the diagonal pattern (with the only exception in training of 1.6-2.2cm ground sampling distance). The off-diagonal performance (when there was considerable mismatch between the training and test resolution) was generally lower in performance, illustrating the importance of having a similar resolution during model training and testing.

\subsection{Solar panel detection performance with respect to flight speed}
Apart from resolution, another important controllable variable for drone imagery operation is the speed at which to collect the imagery. The faster the drone flies, the shorter the time span and hence the lower the cost of data collection. However, greater speed may also introduce additional noise into the data due to motion blur \cite{csimcsek2021novel}, impacts on the stability of the flight, reduced exposure time, and forcing the use of a higher ISO. Each of these may lower the performance of solar panel detection. Here we aim to provide a controlled experiment to investigate the change in performance if we fly at different flight speeds and compare how flight speed impacts detection performance.

The UAV we used for these experiments, the DJI mini 2, has 2 operational flight speeds: "Normal" mode and "Sport" mode. Normal mode has an average speed of ~8m\/s and Sport mode has an average speed of ~14\/s. Continuous videos were taken and individual frames were cut from the video and labeled for evaluating the performance change when flight speed changes (models only operate on individual image frames rather than continuous video). 

\subsubsection{Results}
\begin{figure}[h!]
    \centering
    \includegraphics[width=\textwidth]{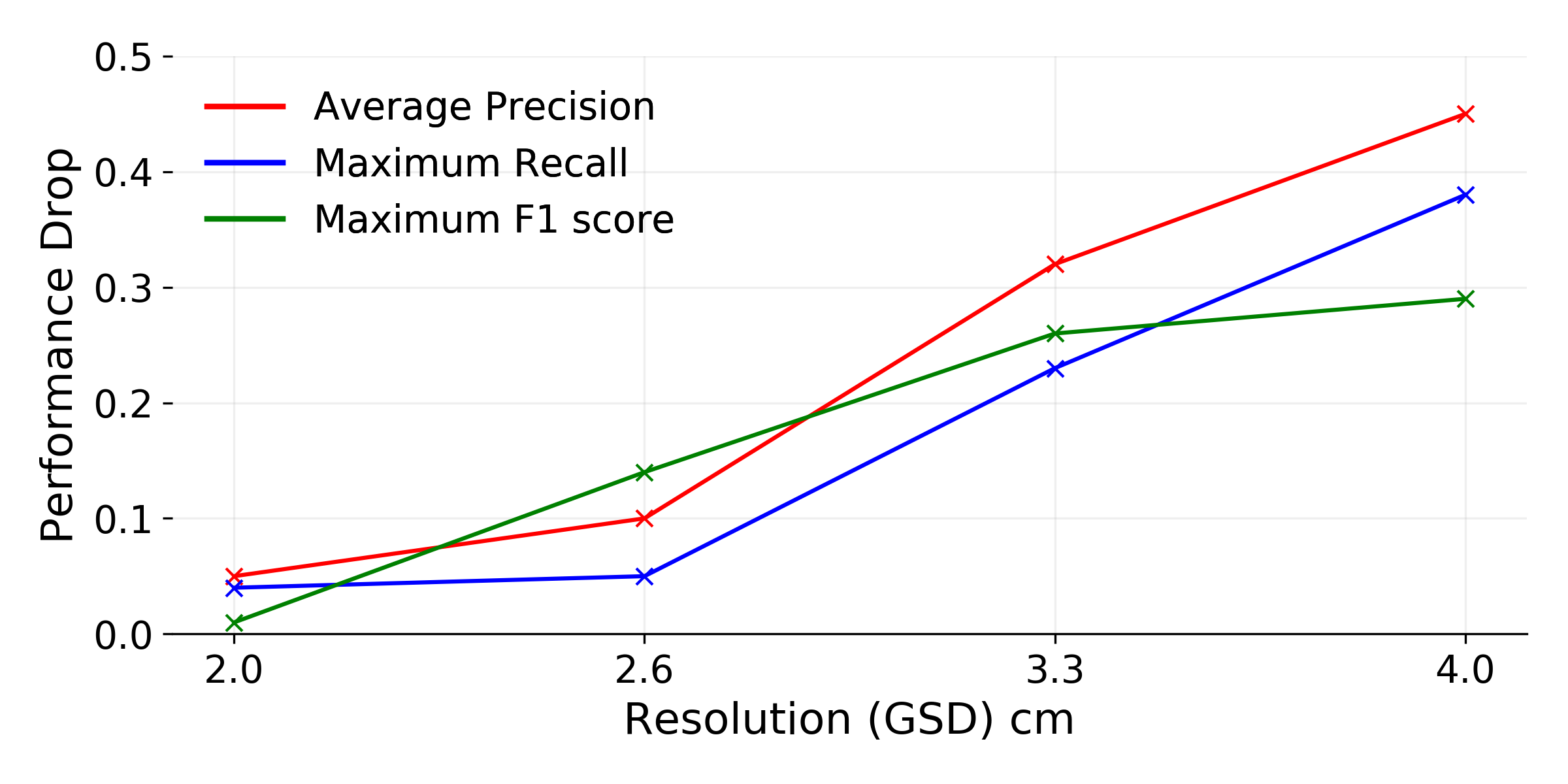}
    \caption{Detection performance drop due to flying faster at various altitude. Formally, it is the absolute difference between a performance metric at slow flying speed and the same performance metric at high flying speed.}
\label{Fig:E2_speed}
\end{figure}

Fig \ref{Fig:E2_speed} shows the degradation in detection performance with respect to flying speed. From the plot we can see that the detection performance does drop with respect to the flying speed as all performance statistics (AP, ${F_1}_{max}$ and $R_{max}$) are worse for sports mode than the normal mode. However, as the resolution differs, the performance drop monotonically increases as the resolution becomes lower. We also note that with the normal mode (slow speed) flying, the detection performance is kept the same comparing to stationary mode.

\section{Experiment \#2: Cost analysis of UAV-based solar panel detection and comparison to satellite data} \label{sec:exp2}

\subsection{Cost analysis: Methods}
Having demonstrated  that UAVs can accurately identify very small solar PV panels under laboratory conditions, a remaining question is whether using UAVs for this purpose is cost-effective. In this section, we estimate the costs of using UAVs for small solar panel detection and compare the estimated costs with that of commercial satellite data and aerial photography. Although we present the operational trade-off of detection performance using small solar PV detection, these costs may be relevant to a variety of similar analyses of optical overhead UAV imagery data.

\begin{figure}[h!]
    \centering
    \includegraphics[width=\textwidth]{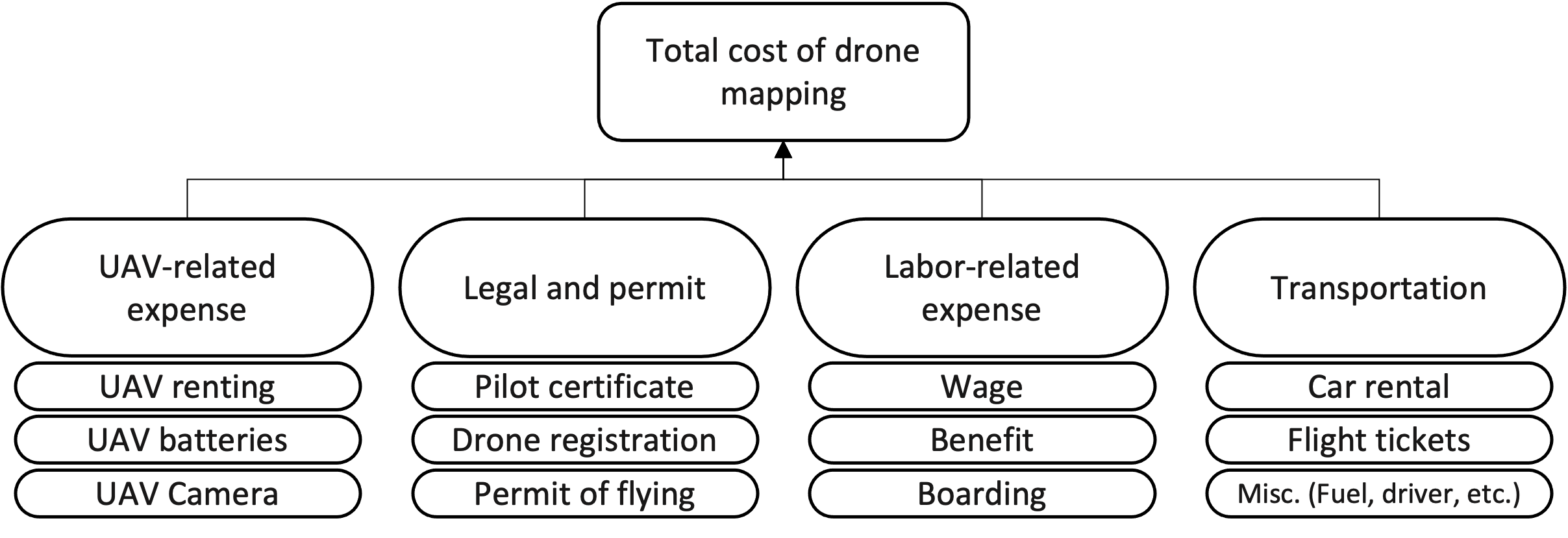}
    \caption{Major categories of our UAV operation cost estimate structure. Legal and permit cost is highly region-dependent and is usually a fixed cost that is not dependent on the resolution and area covered. All other categories are highly resolution-area dependent.}
\label{Fig:Cost_structure}
\end{figure}

To estimate the cost of UAV mapping (in unit of \$), we first identify two key characteristics of the final product: (1) image resolution and (2) size of the area imaged. Nearly all our cost estimates are a function of these two product specifications (e.g., cost per $km^2$ at a resolution of 0.03m). With a fixed resolution and total area, the total amount of work required (total flight time) can be determined. Although we split our cost estimation into five major categories listed in Fig \ref{Fig:Cost_structure}, not all of them are dependent on the image resolution or size (such as legal costs). We made a few assumptions regarding operational parameters in order to provide representative cost estimates for a generic case: 
\begin{itemize}
    \item The UAV is operated five days each week, six hours per day (assuming an eight-hour work day, and allowing two hours for local transportation and drone site setup)
    \item Each UAV operator rents one car and one UAV; the upfront cost of the UAV is amortized over the expected useful life of the UAV
    \item Total UAV image collection time is capped at three months, but multiple pilots (each with their own UAV) may be hired if necessary to complete the collection
    \item UAV lifetime is assumed to be 800 flight hours \footnote{There is no manufacturer suggestion on the lifetime of their machine. A representative at UAV manufacturer that we contacted suggested that 800 flight hours would be a reasonable expectation}
    \item A sufficient quantity of UAV batteries is purchased for operating for a full day
    \item The probability of inclement weather is fixed at 20\% \footnote{This number can vary and sometimes be very high in some seasons within specific places; for example in August in Douala, Cameroon, this number can be larger than 80\% (from \href{https://en.wikipedia.org/wiki/List_of_cities_by_sunshine_duration\#cite_note-CIEH-2}{wiki})} and no operation would be carried out under those conditions.
\end{itemize}
 Using these assumptions, we can translate total required flight time for a given data collection into total working hours, the number of pilots needed and the number of UAVs needed. Together, this information can be combined to estimate each individual cost type we enumerate in Fig \ref{Fig:Cost_structure}.

We distinguish four categories of costs, as illustrated in Fig \ref{Fig:Cost_structure}, for UAV data collection: 
\begin{itemize}
    \item Legal and permit: The legal and permitting cost of getting the credentials for flying in a certain country or region. As an example, in the US, although state laws may vary, at a federal level, flying for non-hobbyist purposes (class G airspace, below 120m) requires the drone pilot to have a Part 107 permit, which requires payment of a fee as well as successful completion of a knowledge test. The legal and permitting costs are inherently location-dependent, and cost variation may be large.
    \item Transportation: The total transportation cost for the drone operator. For the purposes of our estimate here, we assume one drone pilot (thus, total data collection time is a linear function of area covered). Note that this category includes travel to and from the data collection location, which is assumed to include air travel, local car rental, car insurance, fuel costs, and (when the operational crew is foreign to the language) a translation service.
    \item Labor-related expenses: Umbrella category of all labor-related costs including wages and fringe benefits or overhead paid to the drone pilot, as well as boarding and hotel costs.
    \item Drone-related expenses: Umbrella category of all drone-related costs including purchase of the drone, batteries, and camera (if not included with the drone). 
\end{itemize}

The details of the price assumptions are outlined in Table \ref{Tbl:detailed_cost_estimate}. In each row, we specify the major cost item and the unit price. We put additional details in the appendix.

\begin{table*}[h]     
\centering
\begin{tabular}{cccc} 
\toprule
Category                              & Item                    & Unit cost      & Unit       \\ 
\midrule
\multirow{3}{*}{Legal and permit}     & Part 107 certificate    & \$150  \cite{faa_drone_pilot}        & /pilot             \\ \cline{2-4} 
                                      & Pilot training for exam & \$300          & /pilot       \\ \cline{2-4} 
                                      & Drone registration fee  & \$5 \cite{faa_drone_registration} & /drone$\times$year   \\ \hline
\multirow{5}{*}{Transportation}       & Car rental              & \$1,700        & /month    \\ \cline{2-4} 
                                      & Car insurance           & \$400          & /month      \\ \cline{2-4} 
                                      & Fuel                    & \$3 \cite{eia_feul_price} & /gallon            \\ \cline{2-4} 
                                      & Flight ticket           & \$2000         & /pilot   \\ \cline{2-4} 
                                      & Driver/translator       & 0 in US        & /pilot        \\ \hline
\multirow{3}{*}{Labor related}        & Wage                    & \$40           & /hour$\times$pilot     \\ \cline{2-4} 
                                      & Benefit                 & \$20 \cite{bls_compensation}          & /hour$\times$pilot          \\ \cline{2-4} 
                                      & Hotel                   & \$125          & /night     \\ \hline
\multirow{4}{*}{Drone related}        & Drone                   & \$27,000 & /drone       \\ \cline{2-4} 
                                      & Camera                  & \$0   & /drone  \\ \cline{2-4} 
                                      & Battery                 & \$3,000  & /drone      \\ \cline{2-4} 
                                     & Data storage            & \$130          & /5TB     \\ 
\bottomrule
\end{tabular}
\caption{Table for detailed breakdown of the cost assumptions in each categories for US-based operation. References embedded with hyperlinks to click on. Note that nearly all of the unit prices in the above chart vary greatly with the operation location of interest, and therefore anyone referencing this cost estimate framework should adjust according to local conditions and prices. A detailed calculation and explanations about our cost assumptions are given in the appendix and can be found at our code repository. }\label{Tbl:detailed_cost_estimate}
\end{table*}


\subsection{Cost analysis: Result}
\begin{figure}[h!]
    \centering
    \includegraphics[width=\textwidth]{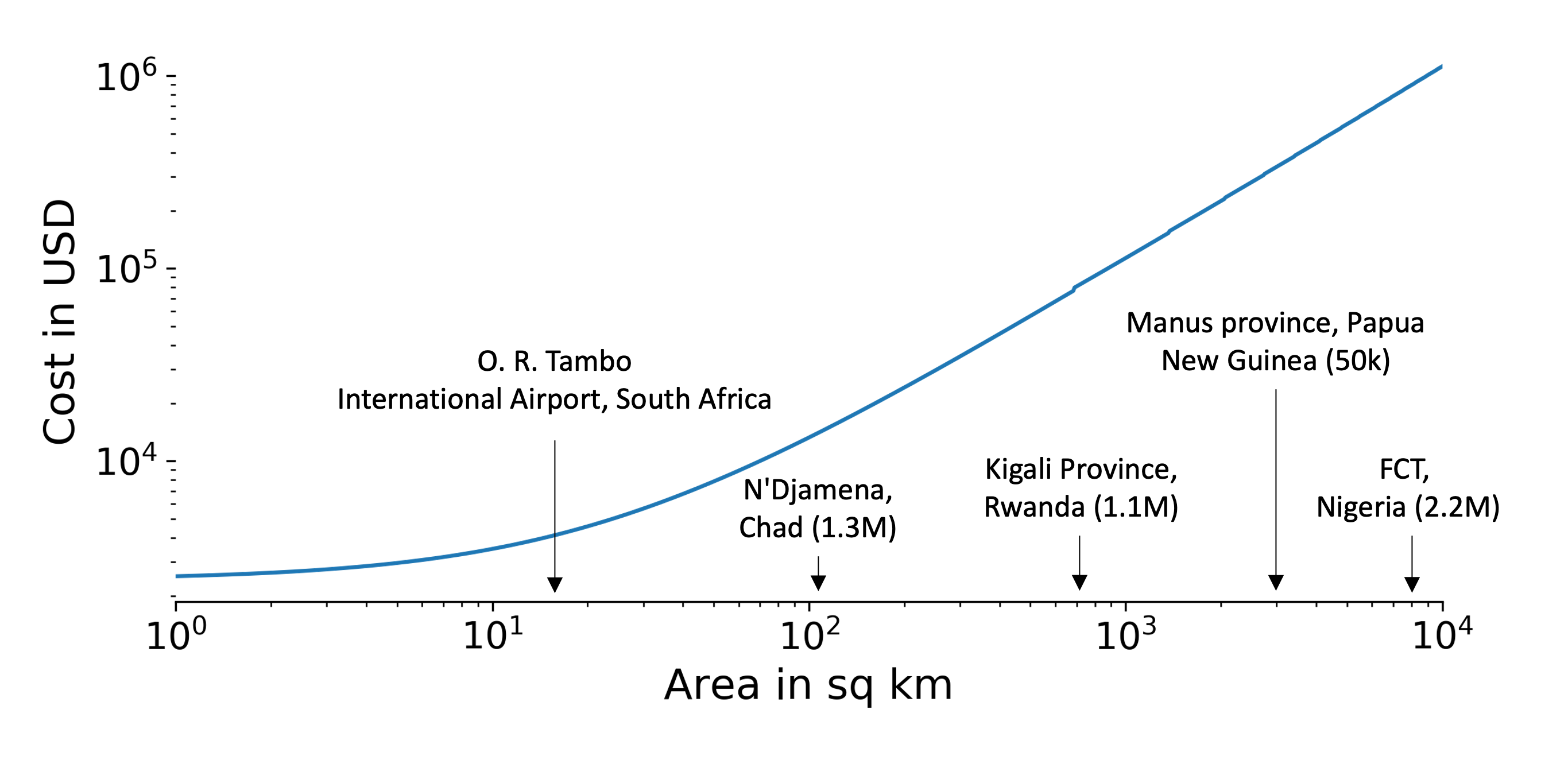}
    \caption{Total cost of UAV mapping with respect to total area mapped with resolution of 0.03m. The x axis is the area in $km^2$ in log scale and y axis is the total cost in USD in log scale. To help make sense of the area, we provide some benchmark area sizes with the population in the area in parenthesis at the end: (1) O.R. Tambo International Airport of South Africa, the busiest airport in Africa; (2) N'Djamena, capital city of Chad; (3) Kigali, capital and largest city of Rwanda; (4) Manus Province, the smallest province of Papua New Guinea; (5) Federal Capital Territory, capital area of Nigeria. All description and population information is extracted from Wikipedia.}
\label{Fig:total_cost_vs_area}
\end{figure}

We present our cost estimate in Fig \ref{Fig:total_cost_vs_area}. There were three types of drones used in \cite{chew2020deep} and we use "Ebee Plus" to estimate our total drone operation cost because it is able to map the largest area per flight (or per unit time) among the three. From Fig \ref{Fig:total_cost_vs_area}, we see that the relationship between the total cost and the total area of interest after some threshold is nearly linear (i.e., as the fixed cost is averaged out over a larger amount of area). We also see that the overall cost of using a UAV to map high-resolution imagery is non-trivial: mapping an area equivalent to the size of the Federal Capital Territory (7,000-8000 $km^2$) around and including Abuja, the capital area of Nigeria, would cost around 6 million dollars in total.

One of our assumptions the price estimation is that the pilots are not within day-trip distance to the place of interest and need lodging during operation. Local pilots are also an option and are increasingly available, which would yield lodging and other travel related expense savings. At 0.03m resolution, the lodging fee, at most represents about 15\% of the total operation cost, which would not impact our conclusion or the relative ranking of UAV, satellite, or aerial imagery cost. More information regarding lodging fee is provided in the appendix.

\begin{figure}[h!]
    \centering
    \includegraphics[width=\textwidth]{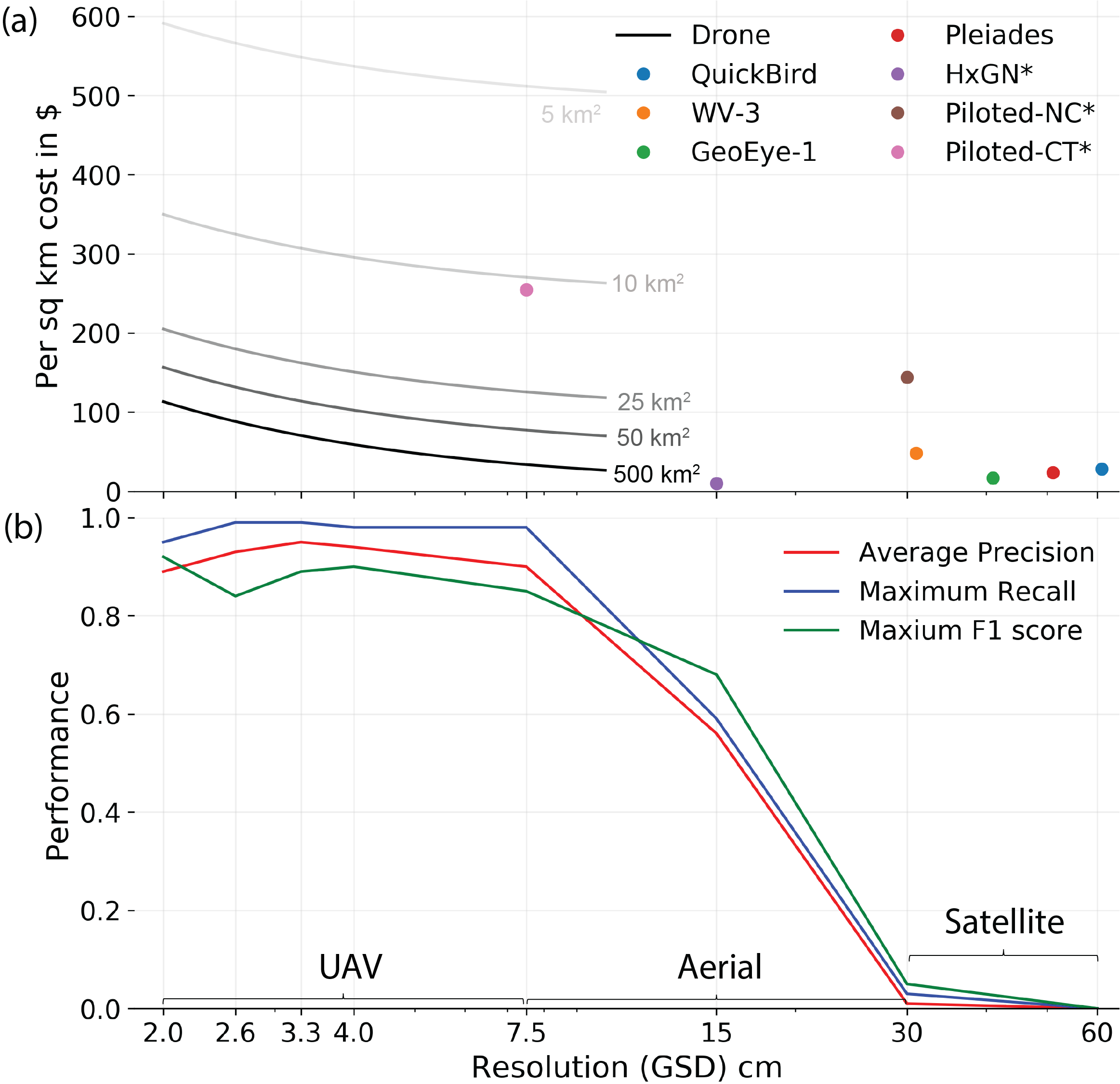}
    \caption{Cost and performance tradeoff with UAV and satellite imagery. (a): Cost per $km^2$ in USD versus resolution for  UAV, satellite, and aerial imagery. Hexagon (HxGN*) provides archival piloted aerial imagery which is limited in coverage compared to satellite coverage. Pilot-\{CT, NC\}* are piloted new aerial imagery collection cost estimated from government documents of CT and NC USA. Note that unit cost varies with different target area size for drone operation. (b): Detection performance vs resolution up to satellite resolution with typical resolution of platform annotated. The performance drops significantly down to effectively zero from typical UAV resolution to typical Satellite resolution}
\label{Fig:cost_compare}
\end{figure}

Apart from the drone operation cost, the core question we aim to answer concerns the cost and performance trade-offs between using drone-based high-resolution imagery or satellite imagery for a given setting or application. Therefore, we combine these two elements together in Fig \ref{Fig:cost_compare} for direct reference. Note that the price estimates for commercial offerings are estimated from online sources, government agency reports, and expert consultation, but we recognize that these estimates; estimates which we believe are reasonable for the purpose of this study but certainly imperfect. The exact price would vary based on negotiated rates, changing business models, and on specific parameters of the desired data collection including the region of interest, business status (e.g. for nonprofit or for-profit use), etc. Currently, the highest resolution (0.3m) commercially available satellite imagery is provided by private companies. We collected estimates of satellite imagery cost per unit area (USD per $km^2$) and compared these with our estimated UAV average cost per $km^2$ with the same resolution. \footnote{We understand the limitation of public quotes and timeliness of these quotes might vary, along with the fact that companies may offer unit pricing discounts when customers purchase relatively large quantities of imagery.} 

From Figure \ref{Fig:cost_compare}, we draw several conclusions: (1) The satellite imagery costs are typically much lower than UAV imagery cost, although they also have significantly lower resolution, which results in extremely poor performance in very small SHS detection. (2) The unit price of UAV imagery is lower both when the total area increases sufficiently (to a limit where the average fixed cost per unit area falls to essentially zero, so that the relationship between total cost and coverage area becomes essentially linear) and when the resolution requirement drops. (3) The cost of UAV imagery is comparable to satellite imagery when operating over a large area (>50 $km^2$) with lower image resolution (4cm-7.5cm); which also achieves excellent performance according to performance plot.

In addition to the cost and performance of satellites and UAVs, Figure \ref{Fig:cost_compare} also provides the same information for one more approach: imagery collection from piloted aircraft. To estimate the costs of the piloted aircraft approach, we collected data from two government reports, \footnote{one from \href{https://files.nc.gov/ncdit/documents/files/OrthoImageryBusinessPlan-NC-20101029.pdf}{North Carolina Geographic Information Coordinating Council} and the other from \href{https://portal.ct.gov/-/media/Office-of-the-Governor/News/2021/20210426-Governor-Lamont-ARPA-allocation-plan.pdf}{Connecticut's Plan for The American Rescue Plan Act of 2021}}, and also information on the Hexagon (HxGN) Imagery Program. That report indicates that the cost is \$100-200 per $km^2$ and the resolution ranges from 7.5cm to 30cm. However, we note that these cost and resolution figures are based on historical data collection in the US, conducted for government agencies and for a large coverage area. Therefore, the average costs may differ in a different setting, such as in South Asia or sub-Saharan Africa. The Hexagon Imagery Program's price reflects only the archival imagery that was collected in previous piloted missions and is limited in geospatial and temporal coverage compared with satellite imagery.

In conclusion, although satellite imagery are cheaper and logistically easier to access, satellite imagery simply cannot adequately capture the small solar PV as Section \ref{sec:exp1} shows. UAV imagery has the advantage of unmatched high resolution and therefore, even at somewhat higher costs, is far more reliable in very small SHS performance in the SHS detection task. While tasks other than small SHS would likely have different performance with respect to image resolution, the trend we show here may be similar for cost estimates relevant to other applications (such as agriculture \cite{chew2020deep}, wildlife conservation \cite{ivosevic2015use}, emergency response \cite{rabta2018drone}, and wildfire monitoring \cite{afghah2019wildfire}, among others). Also, as we found the biggest cost element is the human operator-related cost, which is indispensable currently due to the legal requirement of human-guided operation within the line of sight \cite{floreano2015science}. However, if automated UAV flights become possible (or did not require on-site operators), then the cost of drone-based operations would become significantly lower, potentially giving UAVs a large cost advantage compared to satellites.

\section{Experiment \#3: Case study: Rwanda SHS detection using drone imagery} \label{sec:exp3}
The last of our three questions was to evaluate how well UAV-based SHS detection performs in real-world scenarios. To test the feasibility and robustness of using UAV imagery to detect small SHS in rural areas, we conduct a real-life case study in Rwanda, to detect the SHS that villagers in developing countries are currently using, using the UAV imagery provided by RTI \cite{chew2020deep, RTI_dataset} \footnote{Public access link: https://doi.org/10.34911/rdnt.r4p1fr}. As this imagery was collected from multiple rural agricultural areas in Rwanda representing different agroecological zones without any prior information on the prevalence or locations of SHS, the model performance achieved is expected to be more representative of the actual performance when this method is applied to a new location of interest. The drone was flown at a fixed height yielding a GSD of 0.03m/pixel, which was also covered in the range of our controlled experiments above. We are also making our annotations of these Rwanda data public for reproducibility.
\begin{figure*}[h!]
    \centering
    \includegraphics[width=\textwidth]{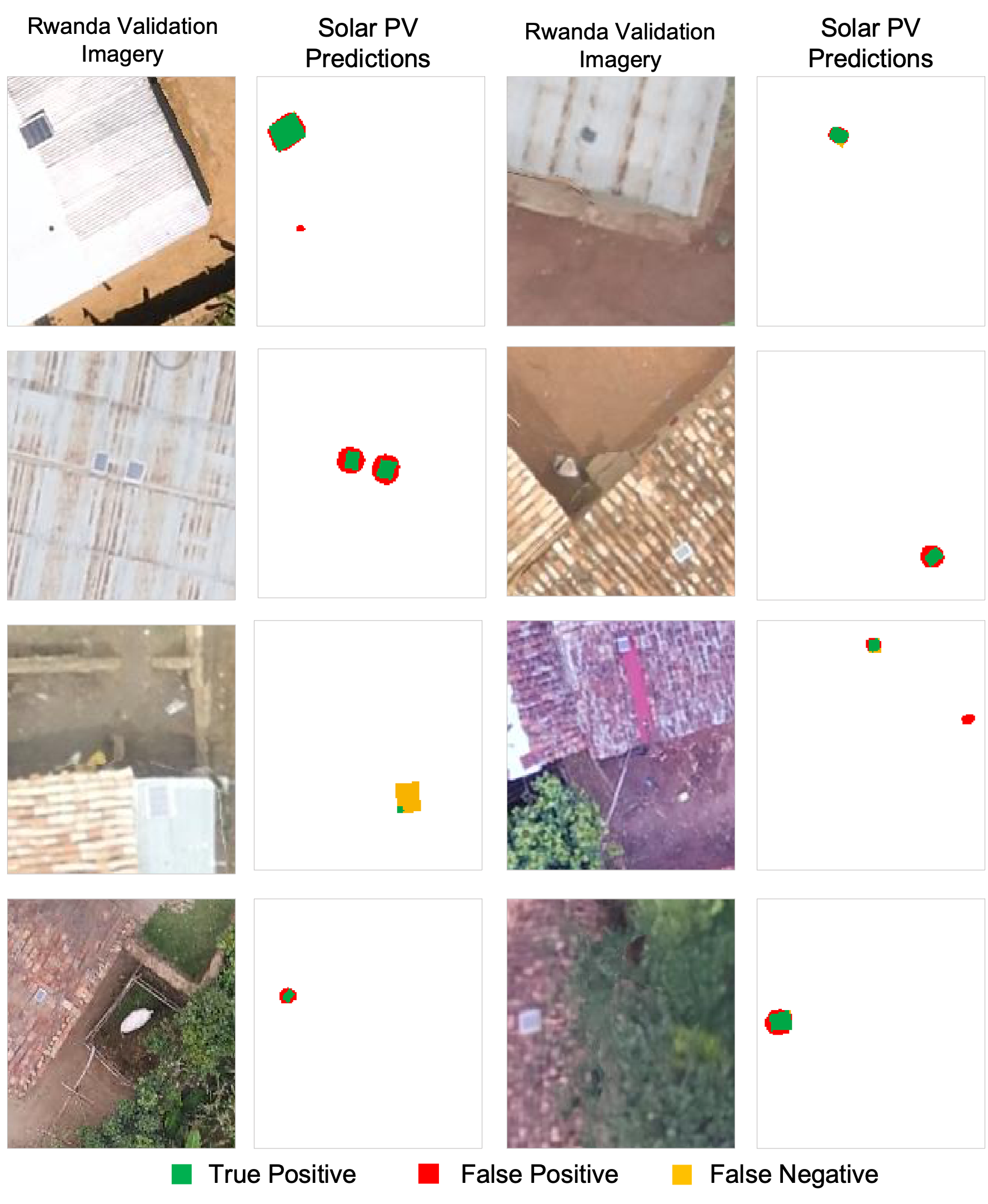}
    \caption{Sample predictions from the Rwanda dataset. Left columns are the imagery patches and right columns show the output come of the predictions. Green represent true positives, which are corrected labelled solar panel pixels. Red represent false positive, which occur when the algorithm predicted a solar panel where there was none. Orange represent false negatives, are actually solar panels, but were not detected.}
\label{Fig:Rwanda_confusion}
\end{figure*}

We manually labeled the solar panels in the Rwanda imagery \footnote{Figshare DOI: \url{10.6084/m9.figshare.18094043}}. We found 214 solar panels in total and split them into a training set of 114 panels and validation set of 100 panels. We limited the fraction of images without SHS to 10\% of the total available to maintain class balance. \footnote{We have around 21k image patches of 512*512 pixels, of which 1\% has solar panels}. In total the model required 2 days to train on the 10\% background data and the 114 solar panels using an NVIDIA GTX2080.

\subsection{Case study: Result}

\begin{figure}[h!]
    \centering
    \includegraphics[width=0.8\textwidth]{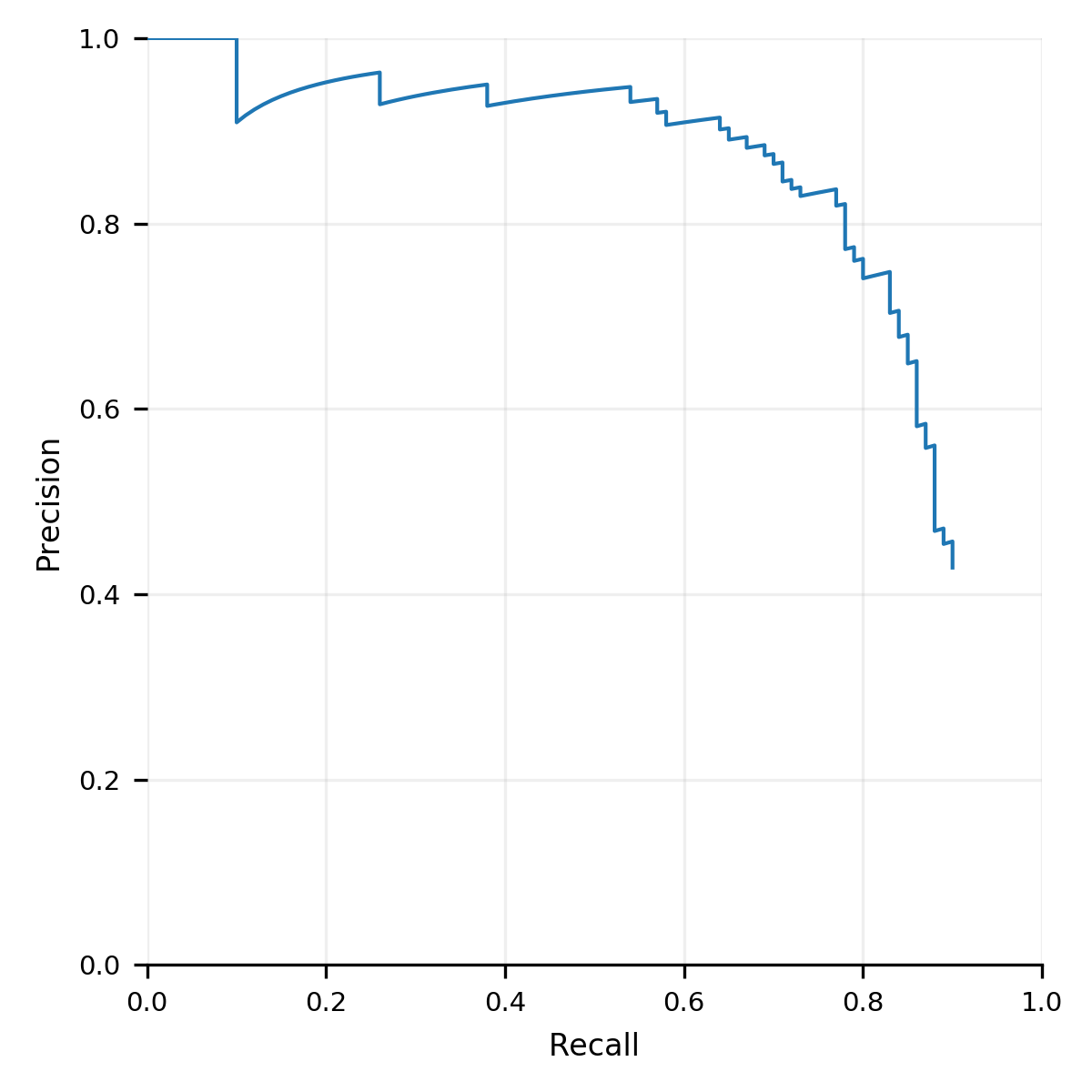}
    \caption{Precision-Recall Curve of the real-life study for small home systems in Rwanda RTI imagery}
\label{Fig:E4_PR}
\end{figure}

Fig \ref{Fig:Rwanda_confusion} shows randomly selected examples of our model predictions. Qualitatively, the majority of the SHS were found with few missing SHS and false positives. Out of the 9 solar panels present in the sample imagery, 8 of them are correctly detected and one of them are missed in the example given. From the example, we can also see that the network can find panels reliably in two distinct roof types present in the Rwanda data.  

Quantitatively, from the PR curve in Fig \ref{Fig:E4_PR} we can see that the performance achieves a maximum $F_1$ score of 0.79, and average precision of 0.8 and maximum recall of around 0.9. By properly thresholding the confidence score, we can achieve a Recall of 0.89 (detecting 89\% of the solar home systems) with precision of 41\%. Lower precision means more false positives, but in a post-processing the detections could be manually reviewed quickly and most eliminated without minimal intervention. Alternatively, recall and be sacrificed for precision, reducing the number of false positives (potentially dramatically so), but sacrificing some recall, meaning that the number of solar panels that were correctly identified would decrease. This amounts to moving around the PR curve in Fig \ref{Fig:E4_PR} to find the appropriate balance for a given application.

Although there is a performance gap between the lab controlled experiment and the Rwanda case study, it is encouraging to see that recall, the fraction of very small SHS successfully identified remains high in both cases, suggesting that UAV based data collection may be a viable approach for providing information to decision makers working in sustainable development.

While these results are encouraging, there remain several limitations or challenges around this application that include regulatory requirements \cite{floreano2015science} to maintain visual line of sight, which necessitates a trained drone operator; limitations in battery technology that bottlenecks the overall flight time of each mission \cite{hassanalian2017classifications}; and limited autonomous navigation and control systems \cite{azar2021drone}. More specifically in sub-Saharan African countries (where the majority of people without access to electricity are located), Washington \cite{washington2018survey} states that the lack of trained operators, lack of government regulations, and privacy concerns are the major challenges to be anticipated for drone technologies to be widely and safely applied to benefit wider communities. However, this has been changing in recent years as more countries in sub-Saharan Africa have developed standard processes for obtaining flight permits and local capacity for operating drones to collect and process imagery for analytical purposes has been expanding rapidly in many countries.


\section{Conclusions}
Small solar home systems such as those commonly used today to provide transitional electricity access to communities are too small for even the highest resolution satellite imagery to capture. We demonstrate that UAV imagery is a viable alternative to map these small solar home systems to provide critical information to stakeholders working to improve electrification in developing countries including the identification of potential markets and information to track progress to electrification sustainable development goals. Through controlled experiments examining the impact of altitude and speed, we tested the technological and financial viability of UAVs for this purpose. We investigated the robustness of drone-based small SHS detection with respect to both resolution (detection performance changes minimally within typical UAV operational altitude) and flying speed (detection performance drops with higher speed), estimated the cost of operation and found comparable cost with satellite imagery given a sufficiently large region to map. We also evaluated UAV small solar detection performance on a case study utilizing UAV imagery in Rwanda that demonstrated that drone-based small SHS detection is a viable approach to supply crucial information about local SHS conditions for energy access projects, successfully identifying nearly 90\% of solar panels with moderate false positive rates that could be reduced through post-processing. 

The evidence from this study suggest that UAVs are a technically viable and financially reasonable approach to collect data for small energy infrastructure like solar home systems. The information about the location and characteristics of small SHS in developing countries collected from drones may provide evidence for decision-making around energy access planning for reaching the SDG 7.1 of universal access to electricity by 2030.

There are at least two potential future directions of this work. The first is developing larger, more diverse collection of UAV data on small energy system objects such as solar PV, diesel generators, electric water pumps for irrigation, and distribution poles and lines. The dataset we share controlled for a large number of possible experimental variables to form fundamental conclusions. A larger and more diverse set of imagery could provide an application-focused benchmark for future algorithm development. The second potential direction of this work is for a larger set of case studies, preferably with even more diverse geographies beyond the regions of Rwanda that we were available for this work to explore algorithm robustness.

\vspace{6pt} 



\authorcontributions{Conceptualization, Jordan Malof, T. Robert Fetter and Kyle Bradbury; methodology,  Jordan Malof, T. Robert Fetter, Robert Beach, Jay Rineer and Kyle Bradbury; software, Simiao Ren; validation, T. Robert Fetter, Robert Beach and Jay Rineer;  investigation, Simiao Ren; resources,Jordan Malof and Kyle Bradbury; data curation, Simiao Ren; writing---original draft preparation, Simiao Ren; writing---review and editing, Simiao Ren, Jordan Malof, T. Robert Fetter, Robert Beach, Jay Rineer and Kyle Bradbury; visualization, Simiao Ren; supervision,Jordan Malof and Kyle Bradbury; project administration, Kyle Bradbury; funding acquisition, T. Robert Fetter and Kyle Bradbury. All authors have read and agreed to the published version of the manuscript.}

\funding{This research was funded by the Catalyst program at the Nicholas Institute for Environmental Policy Solutions, and from the Alfred P. Sloan Foundation Grant G-2020-13922 through the Duke University Energy Data Analytics Ph.D. Student Fellowship. }

\dataavailability{Code for our work is published at:\url{https://github.com/BensonRen/Drone_based_solar_PV_detection}. Two datasets are published to support this work: Main dataset: \url{10.6084/m9.figshare.18093890}; Rwanda PV annotations: \url{10.6084/m9.figshare.18094043}}

\acknowledgments{We would like to express special thanks for Dr. Leslie Collins for providing useful feedbacks and discussion. We also would like to thank Dr. Bohao Huang for his MRS framework code, Mr. Wei (Wayne) Hu for his help in developing the code and discussion, Mr. Trey Gowdy for his helpful discussions and expertise in energy data and other energy fellows / mentors during the Duke University Energy Data Analytics Ph.D. Student Fellowship Program for their suggestions, questions and comments. We also thank the Duke Forest for their use of the UAV flight zone for data collection. }

\conflictsofinterest{The authors declare no conflict of interest}



\abbreviations{Abbreviations}{
The following abbreviations are used in this manuscript:\\

\noindent 
\begin{tabular}{@{}ll}
SDG & Sustainable Development Goal\\
UAV & Unmanned Aerial Vehicles\\
US & United States of America\\
AP & Average Precision\\
IoU & Intersection over Union\\
CNN & Convolutional neural networks\\
SHS & Solar Home Systems\\
GSD & Ground Sampling Distance
\end{tabular}}

\appendixtitles{no} 
\appendixstart
\appendix

\section[\appendixname~\thesection]{}

\subsection[\appendixname~\thesubsection]{Detailed specifications of solar panels used}
\label{App:solarpanel}
The list of solar panels we used is listed in Table \ref{Tbl:solar_panel}. To ensure our experimental results are representative of practical applications of small SHS detection, we chose a diverse set of solar panels from three different manufacturers, included both monocrystalline (black) and polycrystalline (blue) compositions, and varied the physical dimensions in total area and aspect ratio. The majority (4/5) are below 50 Watts and we include one larger panel with a 100 Watt rated power capacity as well (which we imagine to be the upper limit of small SHS).

\begin{table}[h!]

\begin{adjustwidth}{-\extralength}{0cm}
\begin{tabular}{|c|c|c|c|c|c|c|c|c|c|}
\hline
Brand     & X-crystalline  & L (mm) & W (mm) & aspect\_ratio & area ($dm^2$) & T (mm) & Power (W) & Voltage (V) \\
\hline
ECO  & Poly            & 520         & 365        & 1.43          & 19                           & 18        & 25        & 18          \\\hline
ECO   & Mono              & 830         & 340        & 2.45          & 28.3                         & 30        & 50        & 5           \\\hline
Rich solar  & Poly              & 624         & 360.8      & 1.73          & 22.6                         & 25.4      & 30        & 12          \\\hline
Newpowa      & Poly                & 345         & 240        & 1.44          & 8.3                          & 18        & 10        & 12          \\\hline
Newpowa     & Poly              & 910         & 675        & 1.35          & 61.5                         & 30        & 100       & 12    \\\hline
\end{tabular}
\end{adjustwidth}
\caption{Details of the solar panel used in the experiments. X-crystalline (cell composition): Polycrystalline or Monocrystalline, which have slightly different colors in sun-light. L: Length. W: Width. T: Thickness.}
\label{Tbl:solar_panel}
\end{table}

\subsection[\appendixname~\thesubsection]{Satellite view for small SHS}
\begin{figure*}[h]
    \centering
    \includegraphics[width=\textwidth]{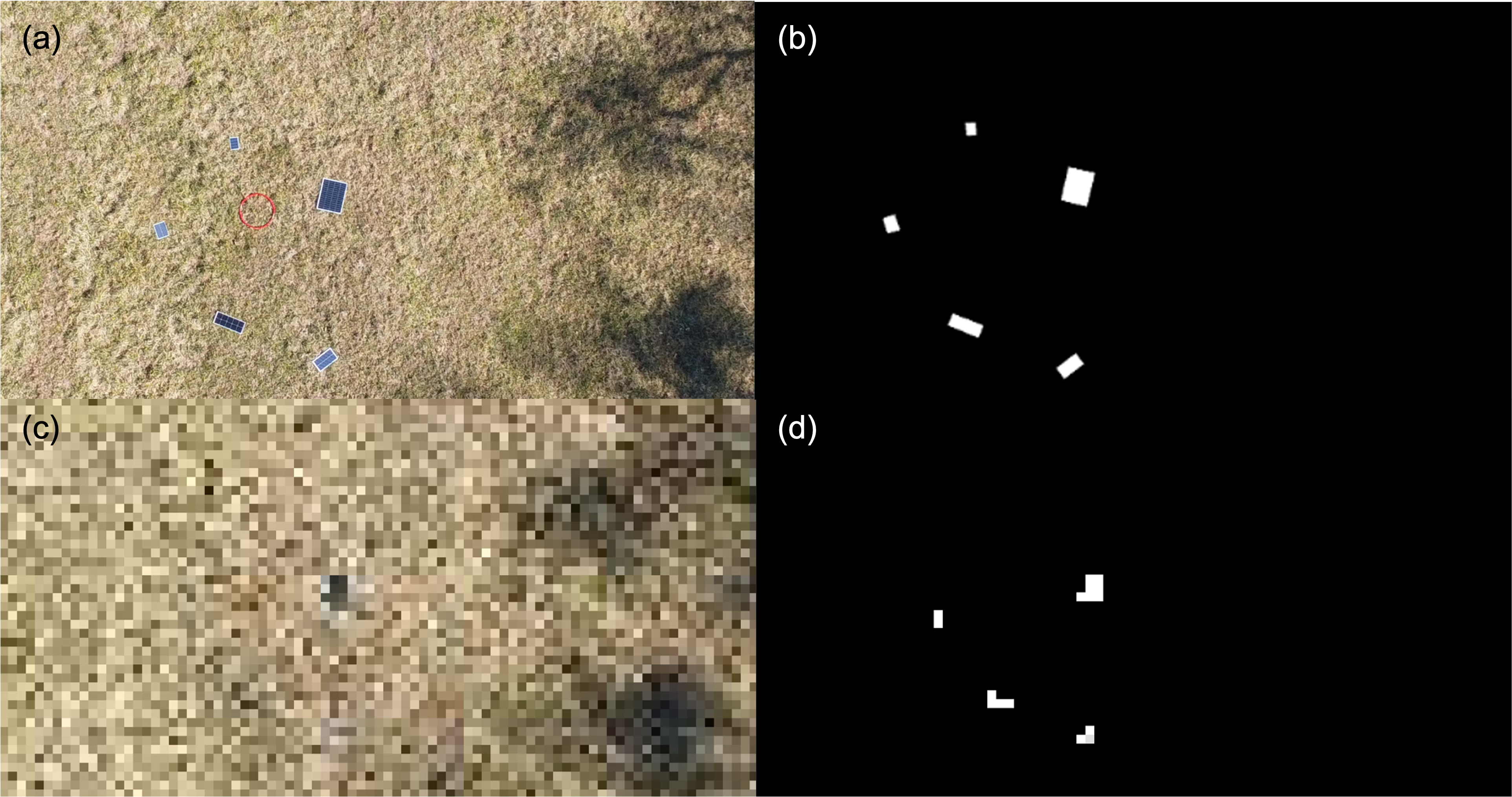}
    \caption{Satellite resolution view of SHS compared with a UAV example. (a) Original UAV imagery with GSD of ~2cm (b) Human labelled ground truth of the SHS (c) Simulated Satellite imagery with GSD=30cm (d) Simulated ground truth of SHS at the resolution of satellite imagery}
\label{Fig:Sat_view_compare}
\end{figure*}

Here we present a simulated (via downsampling) satellite-view of the SHS we investigated in this work to demonstrate the difference in the quality of visibility of the solar arrays at a satellite imagery resolution as compared to UAV image resolution. From Fig \ref{Fig:Sat_view_compare} we see that it is almost impossible (for average human eyes) to detect the presence of solar panels in the image at satellite resolution. We see that the near-zero performance in section \ref{sec:simulated_sate} is reasonable at satellite imagery resolution as there is very limited visibility of these very small SHS. Therefore, UAV imagery resolution is necessary to successfully extract information on the location and size of small SHS.

\subsection[\appendixname~\thesubsection]{Algorithm and performance details}
\begin{figure}[h!]
    \centering
    \includegraphics[width=0.8\textwidth]{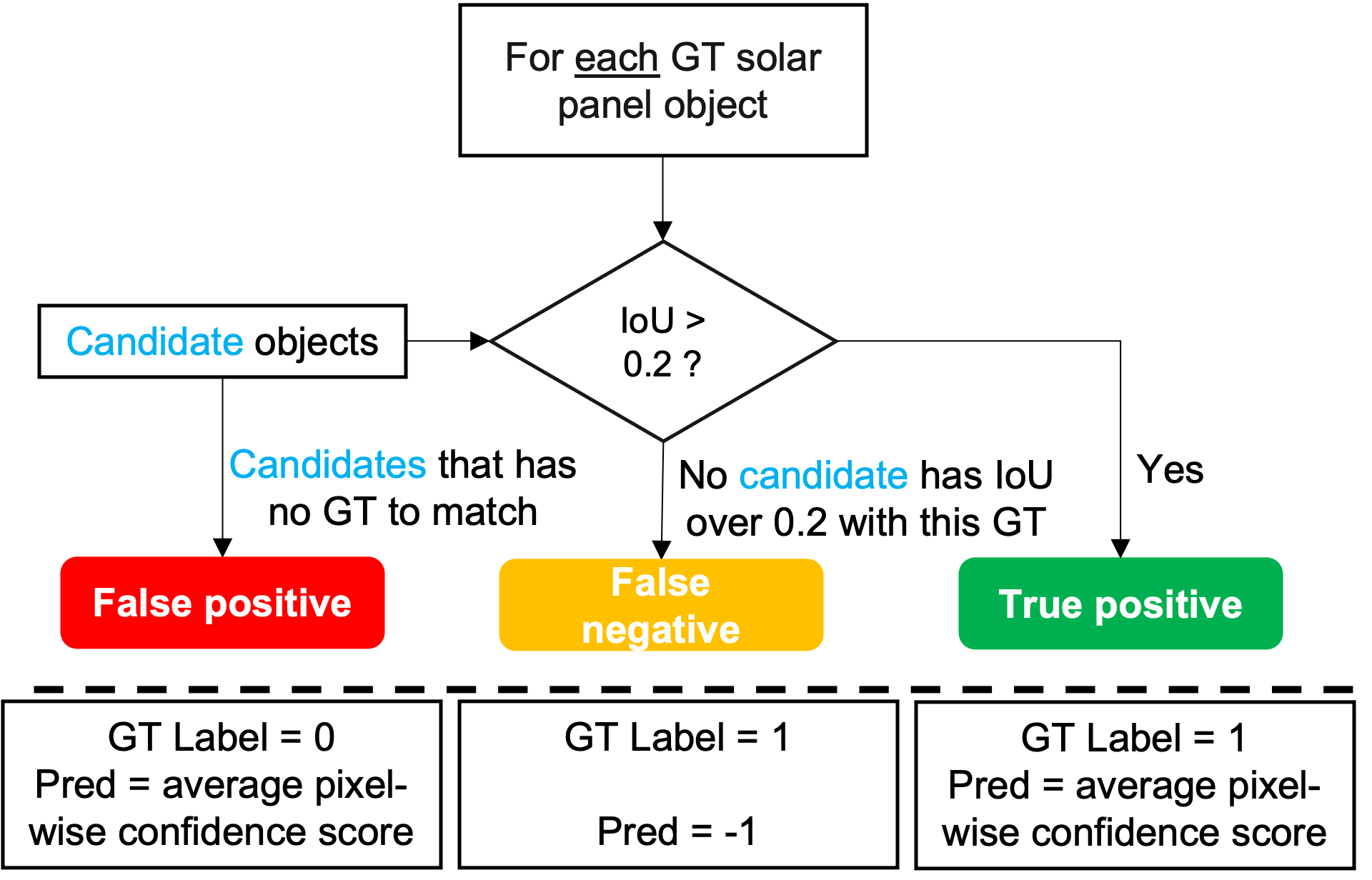}
    \caption{Scoring process. (GT: ground truth). After post-processing, the candidate object groups are matched with the ground truth labels. The confidence score of each object is denoted as the average confidence value of all the pixels associated with that object.}
\label{Fig:Scoring}
\end{figure}

\begin{figure*}[h!]
    \centering
    \includegraphics[width=\textwidth]{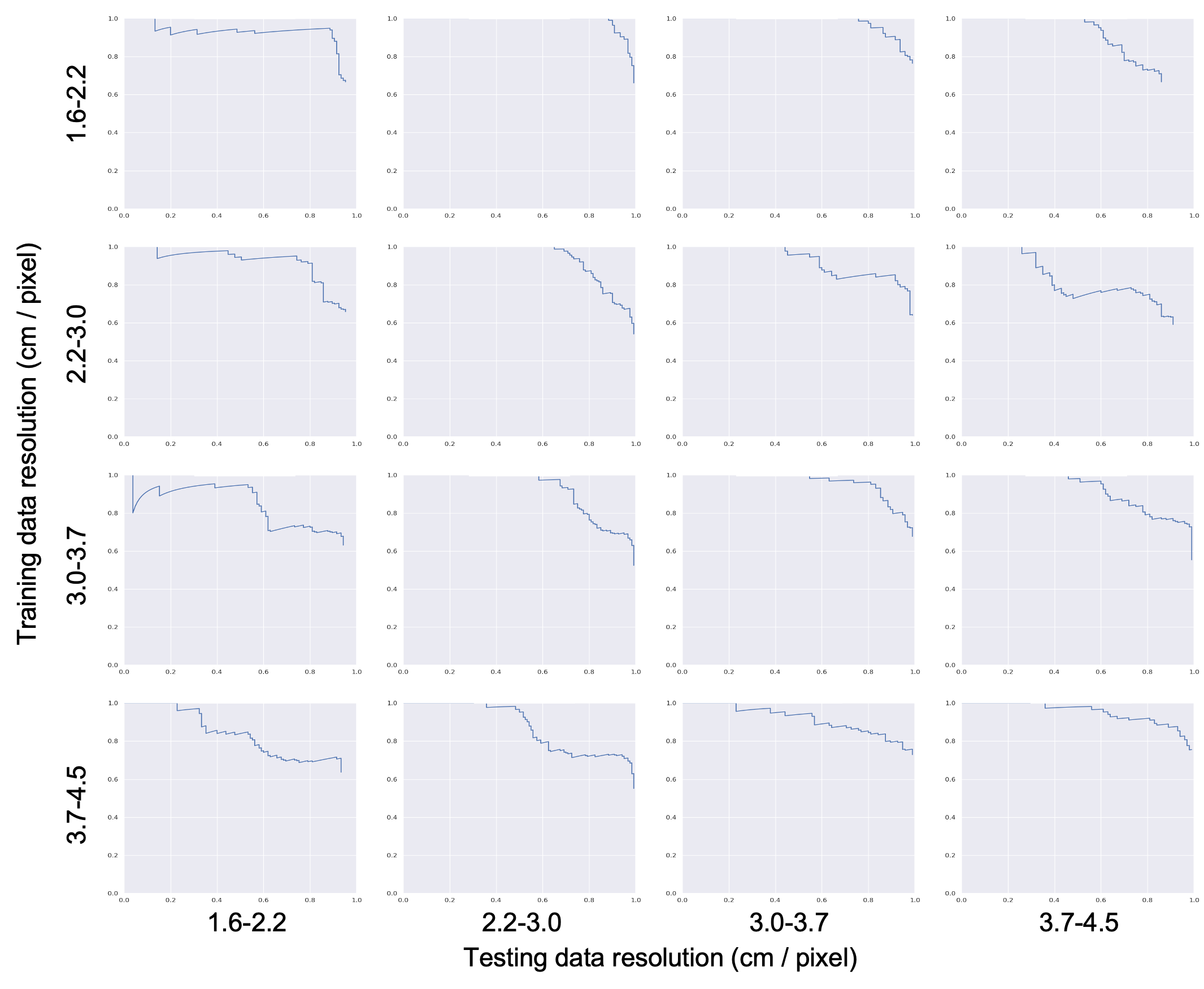}
    \caption{All precision recall curves for the summary metrics shown in Fig \ref{Fig:E1_AP}}
\label{Fig:pair_wise_PR}
\end{figure*}

\begin{enumerate}
    \item \textbf{Pretraining:} As labelled drone datasets, especially the ones including solar panels are extremely scarce, we use satellite imagery containing solar panel (same target as our task, but larger in size) to pre-train our network before fine-tuning it with the UAV imagery data we collected. This practice increased performance over fine-tuning from ImageNet pre-trained weights alone, IoU improved from 48\% to 70\%).
    \item \textbf{Scoring pipeline:} We illustrate the process of scoring in Fig \ref{Fig:Scoring}. Note that in detection problems, the concept of true negatives is not defined. This is also precision and recall (and therefore Precision-Recall curves) are used for performance evaluation rather than ROC curves.
    \item \textbf{Simulating satellite resolution imagery:} In section \ref{sec:exp1}, we downsampled our UAV imagery to simulate satellite imagery resolution. To make sure the imagery has an effective resolution that is the same as satellite imagery, while keeping the same overall image dimensions so that our model has the same number of parameters, we follow the downsampling process with an up-sampling procedure using bi-linear interpolation (using OpenCV's resizing function). The effective resolution remains at the satellite imagery level (~30cm/pixel) but the input size of each image into the convolutional neural network remains the same. \label{apx:resizing}
    \item \textbf{Hyper-parameter tuning:} Across the different resolutions of training data, we kept all hyperparameters constant except for the class weight of the positive class (due to the largely uneven distribution of solar panels and background imagery across changes in GSD). After tuning the other hyperparameters like learning rate and the model architecture once for all flight heights, we tuned the positive class weight individually for each of our image resolution groups due to the inherent difference in the ratio of number of solar panel pixels within each image.
    \item \textbf{Precision-recall curves for Section \ref{Sec:E1_cross_valid}}: As only aggregate statistics were presented in Section \ref{Sec:E1_cross_valid}, we present all relevant precision-recall curves here (Fig \ref{Fig:pair_wise_PR}) for reference.
\end{enumerate}




\subsection[\appendixname~\thesubsection]{Cost estimation calculations and assumptions}
Here we present the details of the calculations and assumptions behind the costs estimates used in this work for UAV operation.

The area that a drone can map in a day $A_{day}$ can be calculated from the manufacture's specifications on the resolution of the imagery of the drone and the battery life for a given flight. We also assume 6 hour daily flight limit on drone operation.
\begin{equation*}
    A_{day} = A_{1flight}(res) / T_{battery} \times 6 (h)\\
\end{equation*}

The area that a drone can map in a single flight is dependent on its resolution and we assume this relationship to be linear (assuming a constant flight speed). The area of the field of view for a drone is quadratic with respect to ground sampling distance. Here we use the manufacture's specifications that claims that flying at a height of 120m produces 3cm ground sampling distance imagery.
\begin{equation*}
    A_{1flight}(res) = res / 0.03 \times 120
\end{equation*}

To calculate the required number of days (workday + weekends) for a specific mission can be calculated using the above quantities. One factor here is that not all days are suitable for a drone flight (e.g. inclement weather). Therefore, we factor in the probability of sunny (or otherwise mild) days into the equation and assume it to be 80\%. This could be replaced with the weather conditions of any region of interest. Also note that since we will use this to calculate the cost of compensation for the pilot, we assume payment during weekends as well although pilots are not assumed to be working during weekends and therefore multiply the number of days required by 7/5 to get the actual time it takes. Note that we assume only weekends (no holiday) for simplicity.
\begin{equation*}
    Day_{tot} = A_{tot} / A_{day} / 5 \times 7 / \%_{sunny}
\end{equation*}

Next we calculate the number of pilots required for the mission, which can be calculated from our assumption that the mission should take less than 3 months (90 days). Since we are paying the pilots by their time, the overall cost would not be largely affected if we increase the number of pilots, hence reducing time for each of them on this mission.
\begin{equation*}
    N_{pilots} = ceil(Day_{tot} / 90)
\end{equation*}

Using the total number of pilots, the fixed cost for the mission can be calculated. The fixed cost consists of pilot training cost, pilot certificate exam cost, permit/registration of drone cost, and transportation (assuming airfare) to the mission site. The flight cost assumption was estimated according to international travel standard ticket fares and assumed to cover up to two round trips based on the assumption of up to 3 month work time. The training cost were estimated from the costs of online preparatory courses for a part 107 certificate.
\begin{align*}
    \$_{fixed} +  &= (\$_{flight} + \$_{training} + \$_{exam} + \$_{permit}) \times N_{pilots} \\
\end{align*}

The bulk part of the operational costs comes from the human labor, hotel, and transportation. Wage information is taken as an average of median wage values reported for drone operators from salary tables across various online sources. The hotel cost assumptions were from average U.S. daily rates from a variety of leading hotel retailers. The car and insurance cost assumptions come from monthly rental quotes from leading car rental companies.
\begin{align*}
    \$_{human} &= (\$_{wage} + \$_{benefit} + \$_{hotel} + \$_{car} \\
    & + \$_{carinsurance} + \$_{translator}) \times Day_{tot} \times N_{pilots}
\end{align*}

Another cost comes from the amortized cost for the uAV. We assume the total flight time for a drone and all associated equipment is 800 flight-hours. The price assumption for drones comes directly from drone retailers. Some drones retailers combine the camera price into the drone price while some others do not. For battery costs, we assume we buy enough batteries to support a full day's operation for each of the pilot teams.
\begin{align*}
    \$_{drone} &= ((\$_{buydrone} + \$_{camera} + \$_{batteries})\\
    &/ T_{lifetime} \times T_{operation}) \times N_{pilot}
\end{align*}

We assume pilots rent cars to commute locally during the mission, therefore, the fuel also needs to be factored into the calculation. Although we found this to be a small enough portion to be neglected retrospectively, we still include it here for completeness. To calculate the total driving distance of the pilot, we assume the mission is to map a large contiguous square area. As the maximum communication distance between drone and controller is 7km, in principle, the pilot needs to drive at least the total area divided by twice the communication distance (14km) to complete the mission. The Mile per Gallon (MPG) of the car is assumed to be 25 MPG and converted to kilometers per Gallon (40 KMPG).

\begin{equation*}
    \$_{fuel} = \$_{gallon} \times A_{tot} / 14 / KMPG
\end{equation*}

Another cost is data storage. Assuming 8-bit color imagery, that translates into 1 Byte / pixel / channel. We assume 4 channels to be stored (red, green, blue, and one for GIS post-processing), the number of pixels can be calculated for a given area and image resolution. The total cost of data storage would be the cost of number of hard drives (HDD) needed to store all the data, which is the total number of bytes divided by the capacity of the HDD (we assume physical hard drives will be needed in case access to cloud services are unavailable).
\begin{align*}
      \$_{DataStorage} &= 1 (Byte / pixel / channel) \times 4 (channels)\\
      & \times area (km^2) / res^2 (m^2) \times 1e6 \\
      & \times\$_{HDD} / Cap_{HDD}
\end{align*}
 
 The total cost of the mission would calculated by adding all of the above costs together:
 \begin{align*}
     \$_{tot} &= \$_{fixed} + \$_{human} + \$_{drone} + \$_{fuel} + \$_{DataStorage}
 \end{align*}

\subsection[\appendixname~\thesubsection]{Lodging cost ratio}
One of our assumptions is the absence of local pilots that does not require a travel status (hotel cost). We recognize this might not be as justified as more and more certified pilots are available across the globe and therefore provide the ratio of lodging fee to the total cost with respect to total area change. 

\begin{figure}[h]
    \centering
    \includegraphics[width=\textwidth]{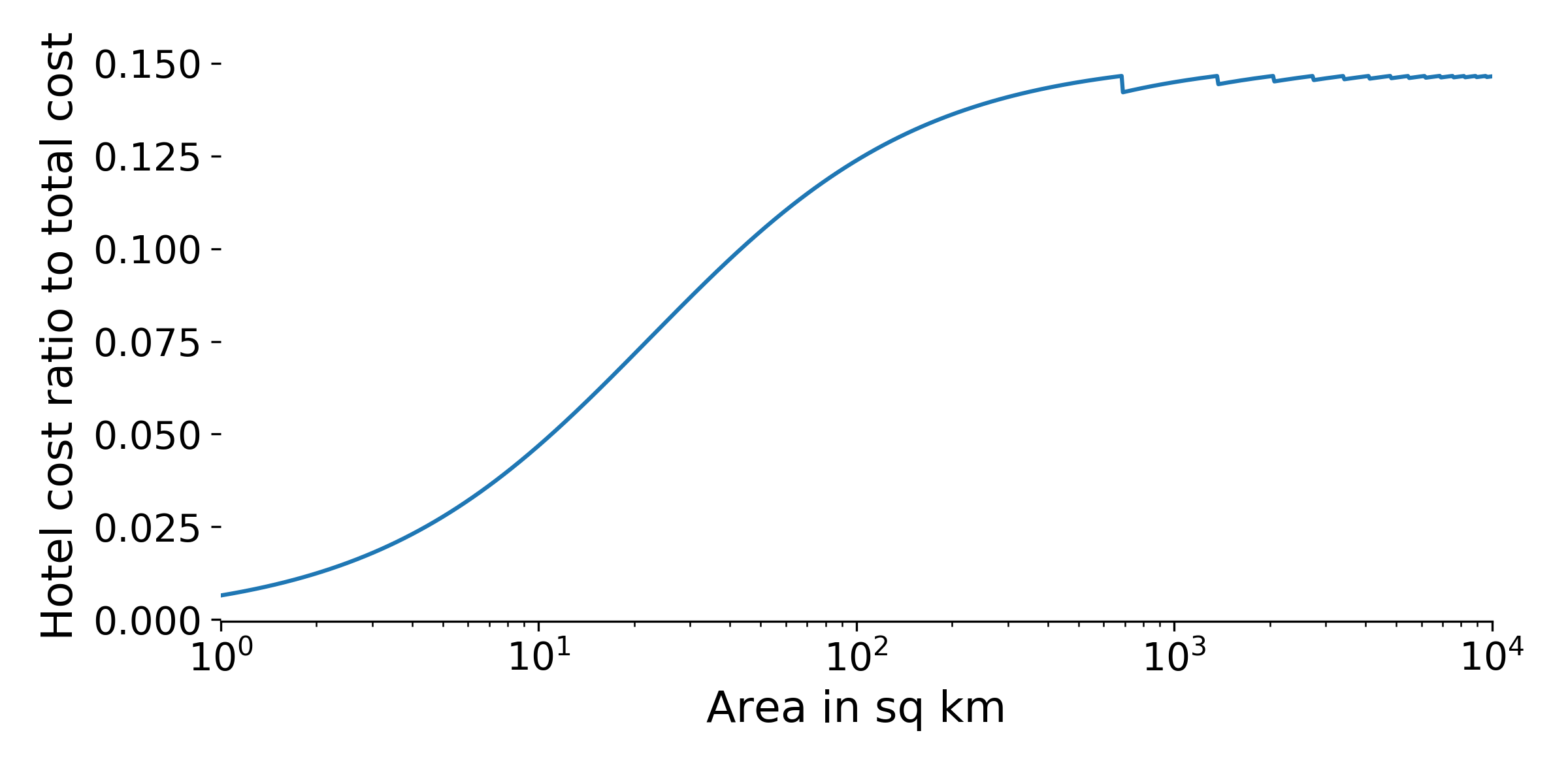}
    \caption{Hotel cost as a percentage of total cost with respect to the total area of the mission at 0.03m resolution.}
\label{Fig:hotel_cost}
\end{figure}

As Figure \ref{Fig:hotel_cost} shows, the hotel cost gradually ramp up when total mission area increase and asymptotes to 15\%. This means that at resolution of 0.03m, if local pilots can be found within day-trip of operation place of interest, up to 15\% of the total cost can be saved.



\begin{adjustwidth}{-\extralength}{0cm}

\reftitle{References}


\bibliography{reference}

\begin{thebibliography}{999}
\bibitem[sol()]{solar_panel_size}
What is the standard size of a solar panel?
\end{thebibliography}

\end{adjustwidth}
\end{document}